\documentclass[twocolumn]{aastex62}

\newcommand\nustar{{\it NuSTAR}~}

\shorttitle{{\it NuSTAR} detection of V5855 Sgr}
\shortauthors{Nelson et al.}
\begin{document}

\title{\nustar Detection of X-rays Concurrent with Gamma Rays in the Nova V5855 Sgr}

\correspondingauthor{Koji Mukai}
\email{Koji.Mukai@nasa.gov}

\author{Thomas Nelson}
\affil{Department of Physics and Astronomy, University of Pittsburgh, Pittsburgh, PA 15260 USA}

\author{Koji Mukai}
\affil{CRESST II and X-ray Astrophysics Laboratory, NASA/GSFC, Greenbelt,
       MD 20771, USA}
\affil{Department of Physics, University of Maryland,
       Baltimore County, 1000 Hilltop Circle, Baltimore, MD 21250, USA}

\author{Kwan-Lok Li}
\affil{Center for Data Intensive and Time Domain Astronomy,
       Department of Physics and Astronomy, Michigan State University,
       East Lansing, MI 48824, USA}
\affil{Department of Physics, UNIST, Ulsan 44919, Republic of Korea}

\author{Indrek Vurm}
\affil{Tartu Observatory, Tartu University, 61602 T\={o}ravere, Tartumaa,
       Estonia}

\author{Brian D. Metzger}
\affil{Department of Physics, Columbia University, New York, NY 10027, USA}

\author{Laura Chomiuk}
\affil{Center for Data Intensive and Time Domain Astronomy,
       Department of Physics and Astronomy, Michigan State University,
       East Lansing, MI 48824, USA}

\author{J.~L. Sokoloski}
\affil{Department of Astronomy, Columbia University, New York, NY 10027, USA}
\affil{LSST Corporation, 933 N. Cherry Ave, Tucson, AZ 85721, USA}

\author{Justin D. Linford}
\affil{Department of Physics, The George Washington University, Washington,
       DC 20052, USA}
\affil{Astronomy, Physics, and Statistics Institute of Sciences (APSIS),
       725 21st St NW, Washington, DC 20052, USA}
\affil{Department of Physics and Astronomy, West Virginia University,
       P.O. Box 6315, Morgantown, WV 26506, USA}
\affil{Center for Gravitational Waves and Cosmology, West Virginia University,
       Chestnut Ridge Research Building, Morgantown, WV, 26505, USA}

\author{Terry Bohlsen}
\affil{Mirranook Observatory, Boorolong Rd, Armidale, NSW, 2350, Australia}

\author{Paul Luckas}
\affil{International Centre for Radio Astronomy Research, The University of Western Australia, 35 Stirling Hwy Crawley, Western Australia 6009, Australia}

\begin{abstract}
We report the first detection of hard ($>$10 keV) X-ray emission simultaneous with gamma rays in a nova eruption.  Observations of the nova V5855 Sgr carried out with the \nustar satellite on Day 12 of the eruption revealed faint, highly absorbed thermal X-rays. The extreme equivalent hydrogen column density towards the X-ray emitting region ($\sim$3 $\times$ 10$^{24}$ cm$^{-2}$) indicates that the shock producing the X-rays was deeply embedded within the nova ejecta.  The slope of the X-ray spectrum favors a thermal origin for the bulk of the emission, and the constraints of the temperature in the shocked region suggest a shock velocity compatible with the ejecta velocities inferred from optical spectroscopy. While we do not claim the detection of non-thermal X-rays, the data do not allow us to rule out an additional, fainter component dominating at energy above 20 keV, for which we obtained upper limits. The inferred luminosity of the thermal X-rays is too low to be consistent with the gamma-ray luminosities if both are powered by the same shock under standard assumptions regarding the efficiency of non-thermal particle acceleration and the temperature distribution of the shocked gas.
\end{abstract}

%% Keywords should appear after the \end{abstract} command. 
%% See the online documentation for the full list of available subject
%% keywords and the rules for their use.

\section{Introduction} \label{sec:intro}

The discovery that nova eruptions are capable of producing transient gamma-ray emission was one of the more surprising results of the {\it Fermi} mission.  The result of a thermonuclear runaway on the surface of an accreting white dwarf, novae are the most common class of stellar explosion known, but far from the most energetic.  Nova eruptions typically eject 10$^{-6}$ to 10$^{-4}$ M$_{\odot}$ of material at velocities of a few 1000 km s$^{-1}$; values previously thought to be too low to lead to efficient particle acceleration.  Yet a growing number of novae have been detected as {\it Fermi}-LAT transients, with MeV to GeV emission lasting for up to two weeks \citep{Ackermann14, Cheung16}. 

Gamma-ray emission in novae is a two-stage process. First, there must be a powerful shock capable of accelerating particles to GeV energies. However, the majority of shock power is expected to remain with the thermal particles, which will be a source of X-ray emission for the likely range of shock velocities (of order 1000 km s$^{-1}$). Gamma-rays are then generated when either the ambient optical photons are inverse Compton scattered by accelerated electrons (the leptonic scenario), or when the accelerated protons collide with ambient matter and produce pions, in particular neutral pions that decay into gamma-rays (the hadronic scenario). In either scenario, gamma-ray emission is expected to be accompanied by non-thermal X-ray emission as well \citep{Vurm18}. These considerations motivate us to search for X-rays in novae concurrent with gamma-ray detection, to better understand the physics of shocks and particle acceleration in novae.

In some of the gamma-ray detected novae, there is evidence that the donor star is evolved. The wind of the donor results in a dense environment into which the nova ejecta expand at high speed, and gamma rays are produced in the forward shock region being driven into the companion wind.  However, most of the {\it Fermi} novae appear to be the more common ``classical'' variety; the donor is typically a lower mass main sequence star without a strong wind, and the binary environment is thought to be ``clean,'' lacking a reservoir of particles to accelerate in the forward shock.  In classical novae, the gamma rays are therefore believed to be accelerated in some internal shock within the ejecta, produced as phases of mass loss with differing velocities catch up and interact with each other.  Such internal shocks have been known to exist in novae for some time, revealed by the presence of faint, hard (1--10 keV) X-rays as observed by {\it Swift} and other X-ray observatories, usually weeks to months after the optical peak. The evolution of the shock X-rays is perhaps best studied in V382~Vel \citep{Mukai01,Orio01b}, while \citet{Orio01a} and \citet{Mukai08} both presented shock X-ray data from many novae. \citet{Metzger14} developed a theoretical framework to explain the observed gamma-ray (and other multiwavelength) emission from novae that assumes internal shocks are present from early in the eruption.

X-ray observations within one or a few weeks of optical peak generally result in non-detection, with a few notable exceptions such as the {\sl ROSAT\/} detection of V838~Her 5 days after optical peak \citep{Lloyd92}. Consequently, no X-ray emission had been detected concurrently with the transient gamma rays until now, making it difficult to quantify the properties of the internal shock or indeed to verify the internal shock picture at all.  One challenge is that any X-ray emission produced by an internal shock shortly after the eruption begins could be embedded deep within the ejecta, and therefore too highly absorbed to detected with the 0.3--10 keV detectors on Swift and other X-ray satellites.  The unprecedented sensitivity of \nustar above 10 keV makes it the ideal instrument to search for this prompt and highly absorbed emission, and efforts have been made to do so. The symbiotic recurrent nova V745 Sco was detected with NuSTAR \citep{Orio15}. However, this detection is not relevant in our context, because this is an embedded nova with an external shock between the nova ejecta and the red giant wind, and because the {\it NuSTAR} observation was carried out well after the putative gamma-ray emission had ended. The observed X-rays were consistent with a thermal origin, and did not reveal new information about the gamma-ray production mechanism.  Two more {\it Fermi}-detected novae, V339 Del and V5668 Sgr, were not detected in \nustar\ observations lasting 23 and 52 ks, respectively (Mukai et al., in prep)\footnote{After the bulk of this work has been completed, another {\sl Fermi}-detected nova, ASASSN-18fv (Nova Carina 2018) was detected with \nustar \citep{Nelson18}.}.

V5855 Sgr (TCP J18102829-2729590) was first discovered by K. Itagaki (CBET 4332) on 2016 Oct 20 (MJD 57681.383), which we take as the start of the eruption (day 0). Optical photometry for V5855 Sgr is presented in a compilation of recent novae by \citet{Munari17}, and we show the V band data in the upper panel of Figure 1 for reference.  The nova eruption was discovered during the initial rise, and its brightness increased from the discovery magnitude of 10.7 (unfiltered) to a peak of $\sim$7.5 four days later.  The nova then had an extended maximum during which the V band magnitude first declined by $\sim$0.5 mag over two days, and then re-brightened over the subsequent week, reaching V $\sim$ 7.5 again on by MJD 57693.5 (Day 12).  After this second maximum, the nova began a secular decline with $t_{2}$ (the time to decay by two magnitudes) of about a week.  \citet{Munari17} note that the first rise to maximum showed a strong wavelength dependence, consistent with the early ``fireball" phase when the ejecta expand in size and drop in temperature.  The second peak is different, showing the same behavior across optical filters.  The authors claim that this second light curve peak is probably connected to the gamma-ray emission.  One possible explanation is that some of the optical light is shock-powered; this production mechanism was proposed to explain the strong optical-gamma-ray correlation observed in the nova ASASSN-16ma \citep{Li17}.

\begin{figure}[htb!]
\begin{center}
\includegraphics[width=3.5in]{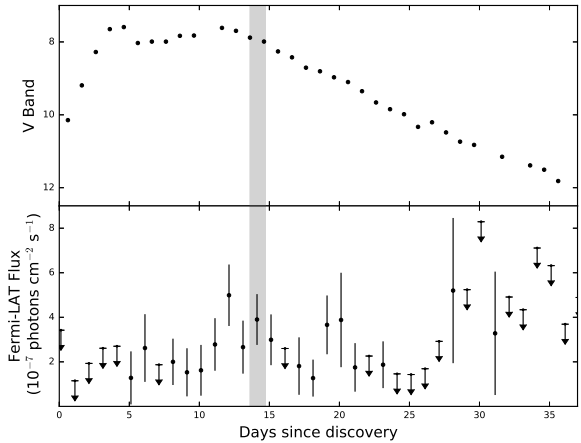}
\caption{{\it Upper panel:} V band light curve of V5855 Sgr from \citet{Munari17}. {\it Lower panel:} {\it Fermi}-LAT daily light curve of V5855 Sgr.  The light grey band indicates the time of the {\it NuSTAR} observation.}
\label{mwl_lcs}
\end{center}
\end{figure}

A number of optical spectra were obtained by amateur astronomers over the course of the eruption: these can be examined at the ARAS\footnote{\url{http://www.astrosurf.com/aras/Aras_DataBase/Novae.htm}} Spectral database, and we present a selection of spectra in Figure 2.  In early spectra P-Cygni profiles were observed in the Balmer lines of hydrogen, with absorption lines extending out to velocities of $\sim$500--900 km s$^{-1}$ \citep{Luckas16}.  After day 12, the continuum became much bluer, the absorption features disappeared and the spectrum was dominated by emission lines.  The lines are broadened and double-peaked, with wings extending out to $\sim$3000 km s$^{-1}$.  The nova became too close to the Sun for optical observations in later November, so unfortunately, no spectra are available after Day 26. 

\begin{figure}[htb!]
\begin{center}
\includegraphics[width=3in]{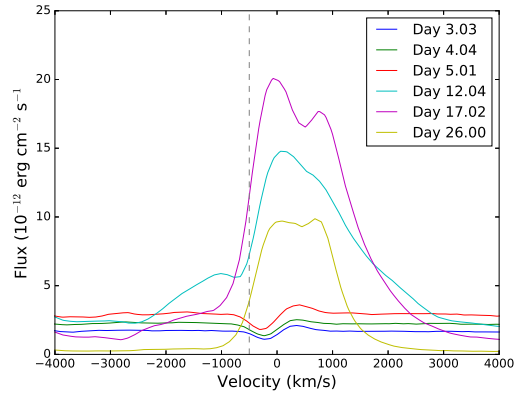}
\caption{ARAS spectra of V5855 Sgr over the first 3 weeks of the eruption showing the evolution of the velocity structure of the H$\alpha$ line.  The dashed grey line indicates the maximum velocity ($\sim$500 km s$^{-1}$) observed in the absorption wing of the early P Cygni profiles.}
\label{mwl_lcs}
\end{center}
\end{figure}

Starting on 2016 October 25 (Day 5 of the eruption), we initiated a target-of-opportunity campaign with the {\it Fermi}-LAT instrument.  V5855 Sgr was detected as a 6-7$\sigma$ source in a binned analysis covering the data range 2016-10-28 to 2016-11-01, with comparable gamma-ray brightness to the systems previously detected by {\it Fermi} \citep[see ][and Section 2, for more detail.]{Li16}.   Based on the gamma-ray detection with {\it Fermi}, we triggered our pre-approved {\it NuSTAR} Cycle 2 target of opportunity program on 2016-11-02.  \nustar was able to begin observing the source within 10 hours of the initial request.  In this paper, we present our analysis of this multiwavelength, target-of-opportunity program, focusing on what we can learn about shocks and gamma-ray production in novae at early times.  We take 2016 Oct 20 (MJD 57681.4) as the start of the eruption, and assume a distance to the nova of 4.5 kpc.  This distance is the value derived from the optical magnitude 15 days after peak \citep{Munari17}.  While the uncertainty in the distance is likely large, most of our conclusions depend on ratios of luminosity and so a precise distance measurement is not necessary.

\section{Observations and Data Reduction} \label{sec:style}

As we discuss in the introduction, we initiated a \textit{Fermi} target-of-opportunity (ToO) campaign of V5855 Sgr shortly after its discovery. The pointed mode used for such target of opportunity programs can increase the exposure time of the target by up to a factor of 2--3 over the standard all-sky scanning mode of \textit{Fermi}. The ToO lasted for the first three weeks of the eruption. We extracted the gamma-ray light curve of V5855 Sgr from the \textit{Fermi}-LAT \texttt{PASS 8} data using the \texttt{Fermi Science Tools (v10r0p5)}. Source events were accumulated within a circular region of interest of radius 20 degrees and an emission model that accounts for Galactic diffuse and isotropic background emission and all nearby sources within 30 degrees from the nova listed in the 3FGL catalog \citep{Acero15} were used. A preliminary daily light curve was first produced using the \texttt{like\_lc} script, which uses the unbinned likelihood method to assess the source significance and flux level for an assumed power-law photon index of $\Gamma_\gamma=-2.3$ \citep[the slope found for nova V959 Mon;][]{Ackermann14}. All other parameters in the emission model were fixed, except for the normalization of the target and the background components.  This light curve then informed more detailed analysis, presented in Section 3.1.  Finally, we ran the \texttt{gtfindsrc} task to constrain the location of the detected gamma-ray source and the nova lies within the 68\% confidence radius of the transient gamma-ray source, confirming their association. 

\begin{figure}[tb!]
\begin{center}
\includegraphics[width=3.5in]{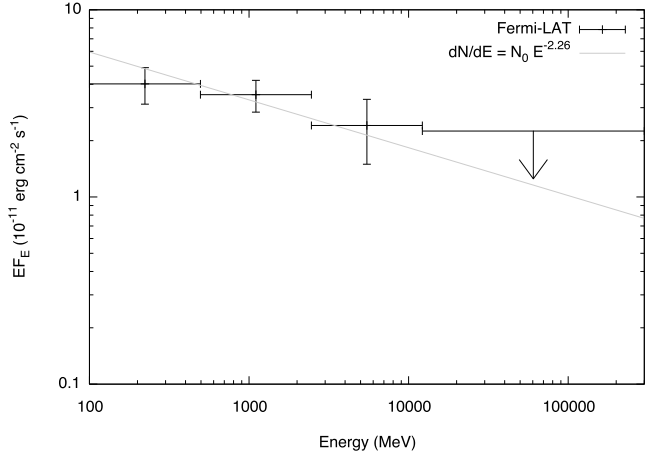}
\caption{{\it Fermi}-LAT spectrum created from data accumulated between MJD 57689 and 57697 (Days 7 and 15 after discovery).}
\end{center}
\end{figure}

{\it NuSTAR} began observing V5855 Sgr on 2016 Nov 02 (Day 13 after discovery).  The observation lasted for 51ks, spread out over $\sim$48 hours.  A small fraction of the observation (about 8ks near the end) was impacted by the drop-out of the {\it NuSTAR} star tracker due to the presence of the crescent moon in the field of view, and we excluded those data from our analysis. To further clean the data of periods of high background due to the satellite traversing the SAA, we re-ran the {\it NuSTAR} data analysis pipeline with strict filtering of SAA passages and removal of the ``tentacle" feature (saamode=strict, tentacle=yes), resulting in a final net exposure time of 40.04 ks.  Images and source spectra were created using the {\tt nuproducts} software for each of the two focal plane modules.  The software produces cleaned images in sky coordinates, and spectra, effective area and response matrix files for specified source and background regions (see Section 3.1 for details of the region files chosen).

\begin{figure*}[tb!]
\begin{center}
\includegraphics[width=5in]{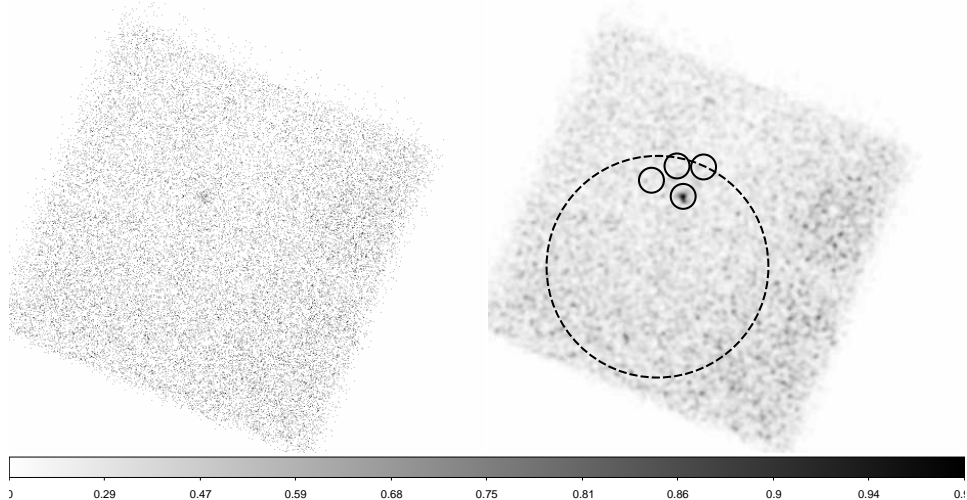}
\caption{{\it Left:} FPMA image with no spatial smoothing.  A faint source is just apparent near the boresight of the telescope. {\it Right:}The same image, smoothed using a Gaussian Kernel of width 4 pixels.  The color bar has had the sinh function in ds9 applied, to make the source more apparent. The solid black circles indicate the regions used to estimate the source and background counts.  The dashed black circle indicates the 68\% positional uncertainty of the {\it Fermi}-LAT source.}
\label{chandra_spec}
\end{center}
\end{figure*}

V5855 Sgr was not detected in two additional X-ray observations carried out with the {\it Swift} satellite on 2016 October 27 and November 6.  The exposure times for the first and second observation were 3.4 and 3.9 ks, respectively, and both were carried out in window timing mode due to the high optical brightness of the source (which can cause pile-up in the CCD in photon counting mode).  The limits on the 0.3-10 keV count rate were $<$0.004 and $<$0.003 counts s$^{-1}$ in the first and second observations, respectively.  

\section{Data Analysis and Key Results} \label{sec:results}
\subsection{{\it Fermi}-LAT detailed analysis}
The gamma-ray emission of V5855 Sgr varied substantially over the first month of the eruption. In the first-pass {\tt like\_lc} light curve, the gamma-ray emission was found to be most prominent between MJD 57689 and 57697 (Days 7 and 15 after discovery).  With the {\it Fermi}-LAT data collected in this interval, we ran a binned likelihood analysis in order to better characterize the gamma-ray spectrum of the nova. In the fitting process, we allowed the normalization of the 3FGL sources closest to the nova (i.e., within 3 degrees) to vary in order to minimize the contamination from them. The best-fit parameters for a single power-law are $\Gamma_\gamma=-2.26\pm0.12$ and $F_{ph}=(2.96\pm0.79)\times10^{-7}$ ph cm$^{-2}$ s$^{-1}$, with a test statistic ($TS$) of 94 (i.e., $\sqrt{TS}$ is approximately equal to the detection significance for the nova). A power-law with an exponential cutoff was also tried, however, without significant improvement. We also did an energy-resolved analysis and the spectral energy distribution (SED) is entirely consistent with the above results (Figure 3).  

In order to facilitate comparison of the {\it Fermi}-LAT spectrum with the theoretical models of \citet{Vurm18}, we derived the monochromatic flux at 100 MeV from the best-fit spectrum, finding $F_{\nu}$ = 2.2 $\times$ 10$^{-33}$ erg s$^{-1}$ cm$^{-2}$ Hz$^{-1}$. This implies $\nu F_{\nu}$ = 5.3 $\times$ 10$^{-11}$ erg s$^{-1}$ cm$^{-2}$.  Finally, with the best-fit emission model in place, we re-ran \texttt{like\_lc} for a daily light curve (parameters of all point sources are fixed, excepted for the normalization of the nova). The results are shown in the lower panel of Figure 1. The integrated flux in the 0.1--300 GeV range during the {\it NuSTAR} observation was 3.1 $\pm$ 0.9 $\times$ 10$^{-10}$ erg cm$^{-2}$ s$^{-1}$, implying a luminosity of 7.1 $\pm$ 2.1 $\times$ 10$^{35}$ ($D$/4.5 kpc) erg s$^{-1}$.

\subsection{{\it NuSTAR} Imaging}
We show the filtered FPMA image output by the nuproducts software in the left panel of Figure 4.  A very faint X-ray source is apparent, even by eye, at the location of the nova in the image (it is more difficult to see in the FPMB image due to the higher background level and the stray light in the upper right of the detector).  The centroid position of the detected X-rays in both modules is offset from the co-ordinates of the optical nova, by 7.1 and 4.8 arcseconds for the A and B modules, respectively.  This offset is well within the published astrometric uncertainty of {\it NuSTAR} (Harrison et al. 2013), and no other X-ray sources are known at this position, so we are confident that the observed X-rays are associated with V5855 Sgr.  To make the nova emission easier to discern, we present the module A image smoothed by a Gaussian kernel of width 3 pixels and shown with sinh-scaling bounded by the minimum and maximum pixel values, in the right panel of Figure 4.

We used aperture photometry to assess the significance of the X-rays at this position. We extracted source counts from a circular region of radius 30 arcseconds centered on the source, and background counts from 3 circular regions, also of 30 arcseconds radius, spaced around the source (see Figure 4, right panel).  Our goal with this choice of background region was to sample, and average out, any gradient in the background light present at this position on the chip.  We detected 202 (366) counts in the source (background) regions in FPMA, and 223 (458) counts from the source (background) regions in FPMB, implying 86 and 70 source counts and detection significances of 8 and 5.7$\sigma$ in the FPMA and FPMB images, respectively.  

\subsection{{\it NuSTAR} spectral analysis}
Although only a small number of net counts remain after background subtraction, we attempted to model the X-ray spectrum using simple models in XSPEC version 12.9.0n.  The source spectra were binned to have a minimum of one count per bin, and model parameters estimated using the modified C-statistic \citep{Cash79} implementation that models the background spectrum bin-by-bin.  The majority of the detected X-ray photons have energies between 10 and 20 keV.  The lack of signal at low energies implies a high degree of foreground absorption, which we include in our models using the photoelectric absorption model {\tt phabs} in XSPEC. For all models, an additional constant is included to account for any calibration offset between the two focal plane module instruments. 

\begin{figure*}[tb!]
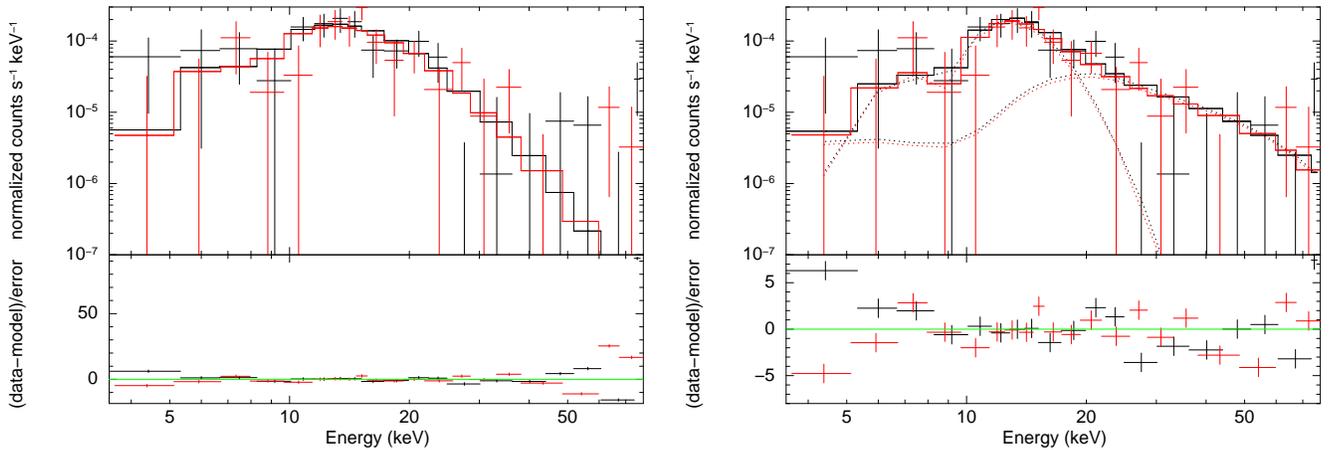

\begin{center}
\includegraphics[height=3.5in, angle=-90]{brems_only.ps}
\includegraphics[height=3.5in, angle=-90]{brems_po1.0.ps}
\caption{{\it Left:} {\it NuSTAR} spectra (FPMA in black, FPMB in red) with best fit absorbed bremsstrahlung model (line 1 of Table 1). {\it Right:}Data with best fit bremsstrahlung plus power law model (line 3, Table 1). The spectral index of the power law is fixed at $-$1.0.  The two components of the model are shown as dotted lines.}
\label{chandra_spec}
\end{center}
\end{figure*}

We first explored two simple models - an absorbed Bremsstrahlung continuum emission from thermal plasma ({\tt phabs*brems}), and an absorbed power law ({\tt phabs*po}). We note that the power law model implemented in XSPEC has the form $f(E) \propto E^{-\alpha}$, where f(E) is the number of photons per energy interval dE, and that we refer to $-\alpha$ (and not $\alpha$) as the photon index in this paper. These models are appropriate for thermal and non-thermal emission from shocked plasma.  The best-fit parameters and associated 1-$\sigma$ uncertainties for each model are shown in Table 1.  Both the thermal plasma and power law models result in a good fit to the data with C-statistic/degrees of freedom of 375.6/380 and 374.9/380, respectively.  The absorbing column ahead of the X-ray emitting region is extremely high, with N(H) = 2.2$^{+0.8}_{-0.5}$ $\times$ 10$^{24}$ and 2.9$^{+1.0}_{-0.8}$ $\times$ 10$^{24}$ cm$^{-2}$ for the bremsstrahlung and power-law models, respectively.  The best-fit plasma temperature is 11$^{+11}_{-5}$ keV (1.3$^{+1.3}_{-0.6} \times 10^{8}$ K).  The unabsorbed flux in the energy range 0.3--78.0 keV is 3.3$^{+6.7}_{-1.5}$ $\times$ 10$^{-12}$ erg s$^{-1}$ cm$^{-2}$ , implying a luminosity of 8$^{+15}_{-1} \times 10^{33}$ erg s$^{-1}$ for our assumed distance to the nova of 4.5 kpc.  The best-fit power law model has $\alpha$ of 3.6$^{+1.3}_{-1.0}$, and the unabsorbed flux is $<$5.3 $\times$ 10$^{-11}$ erg s$^{-1}$ cm$^{-2}$, giving a luminosity $<$3.8 $\times$ 10$^{35}$ ($D$/4.5 kpc) erg s$^{-1}$.  For comparison to theoretical models of non-thermal emission in novae, we also evaluated the monochromatic power law flux $EF_{E}$ at 20 keV, finding (9 $\pm$ 2) $\times$ 10$^{-13}$ erg s$^{-1}$ cm$^{-2}$.

There is some evidence of excess signal at high energies in the residuals of both the thermal plasma and power law model fits (See Figure 5 left).  The low-energy tail of the gamma-ray emission is expected to be detectable at energies above 10 keV as a rising power-law component with $\nu F_{\nu} \propto \nu$ or $\nu^{0.8}$ \citep{Vurm18}, and such a component could exist in tandem with a lower energy thermal plasma.  In order to assess the presence of this non-thermal emission, we added a power law component to the absorbed thermal plasma model and fixed the power law index to be either $-$1 or $-$1.2.  We then obtained the best-fit parameters of the thermal plasma, and both the monochromatic flux at 20 keV and integrated flux of the power law component. The two-component models also fit the data well (C-statistic/degrees of freedom of 371.3/379 and 371.1/379, for power law indices of $-$1.0 and $-$1.2, respectively), and reduce the residuals at high energies (see Figure 5 right).  In both models, the best-fit bremsstrahlung temperature is substantially lower than in the single component model; around 2 keV for both assumed power law indices.  These lower temperature models fall more steeply in the 10--20 keV band, and so larger fluxes are required to produce the observed count rate.  We find best-fit unabsorbed fluxes of  2.7$^{+26.0}_{-2.5}$ $\times$ 10$^{-10}$ (4.3$^{+76.1}_{-4.0}$ $\times$ 10$^{-10}$) erg s$^{-1}$ cm$^{-2}$ for $\alpha_{\rm PL}$ = 1.0 (1.2), implying 0.3--78 keV thermal X-ray luminosities of 6.3$^{+59.5}_{-5.7} \times 10^{35}$  (9.8$^{+174.3}_{-9.5} \times 10^{35}$) ($D$/4.5 kpc) erg s$^{-1}$.  The non-thermal power law component only contributes a small amount of flux in the {\it NuSTAR} band, with inferred 0.3--78 keV fluxes of (2.1 $\pm$ 0.2) $\times$ 10$^{-12}$ erg s$^{-1}$ cm$^{-2}$ for both assumed power law indices.

\begin{deluxetable*}{lcccccccc}
\tablecaption{Best-fit model parameters}
\tablewidth{6.5in}
\tabletypesize{\footnotesize}
\tablehead{Model & N(H)  & kT  & norm\tablenotemark{a} & $\alpha_{\rm PL}$ & $F_{20}$\tablenotemark{b} & $F_{u,0.3-78}$\tablenotemark{c} & C & cstat/dof\\
 & (10$^{24}$ cm$^{-2}$) & (keV) & (10$^{-4}$) & & (erg cm$^{-2}$ s$^{-1}$) & (erg cm$^{-2}$ s$^{-1}$) & & \\}
\startdata
C*phabs*brems & 2.2$^{+0.8}_{-0.5}$ & 11$^{+11}_{-5}$ & 5$^{+9}_{-3}$ & \nodata & \nodata & 3.3$^{+6.7}_{-1.5}$ $\times$ 10$^{-12}$  & 1.0 $\pm$ 0.2 & 375.6/380\\
C*phabs*po & 2.9$^{+1.0}_{-0.8}$ & \nodata & \nodata & 3.6$^{+1.3}_{-1.0}$ & (9 $\pm$ 2) $\times$ 10$^{-13}$ & $<$5.3 $\times$ 10$^{-11}$ & 1.0 $\pm$ 0.2 & 374.9/380\\
C*phabs*(brems+po) & 4.9$^{+1.6}_{-1.7}$ & 2.2$^{+1.7}_{-0.6}$ & 2200$^{+24900}_{-2100}$ & 1.0 & 5.2$^{+2.0}_{-1.8}$ $\times$ 10$^{-13}$ & 2.7$^{+26.0}_{-2.5}$ $\times$ 10$^{-10}$ & 1.0 $\pm$ 0.2 & 371.3/379\\
C*phabs*(brems+po) & 5.1$^{+1.7}_{-1.6}$ & 2.1$^{+1.2}_{-0.6}$ & 3500$^{+44400}_{-3400}$ & 1.2 & 5.8$^{+2.2}_{-2.0}$ $\times$ 10$^{-13}$ & 4.3$^{+76.1}_{-4.0}$ $\times$ 10$^{-10}$ & 1.0 $\pm$ 0.2 & 371.1/379\\
\enddata
\tablenotetext{a}{The normalization of the {\tt brems} model is defined as 3 $\times$ 10$^{-15}$/4$\pi$D$^{2}$ $\int n_{e} n_{H} dV$, where $D$ is the distance to the source in cm, and $n_{e}$ and $n_{H}$ are the densities of elections and hydrogen, respectively, in the shocked plasma.}
\tablenotetext{b}{The monochromatic flux $EF_{E}$ of the power-law component evaluated at 20 keV.}
\tablenotetext{c}{The unabsorbed flux of the bremsstrahlung component in the energy range 0.3--78 keV, evaluated using the {\tt cflux} component in XSPEC.}
\end{deluxetable*}

We summarize our key results as follows: V5855 Sgr is a nova with variable gamma-ray emission over the first month of the eruption.  The bulk of the gamma rays are detected between days 7 and 13, and are best described by a power-law with photon index $\Gamma$ = -2.3.  With our {\it NuSTAR} observation we have detected X-ray emission concurrently with gamma rays for the first time in a nova.  The low signal-to-noise spectrum is equally well described by thermal plasma, power law, or plasma plus power law models absorbed by a very high column of N(H) $>$ 2 $\times$10$^{24}$ cm$^{-2}$.  In the next section, we discuss the implications of these key findings on the nature of the shocks driving the high energy emission in V5855 Sgr.

\section{Discussion}
\subsection{First simultaneous X-ray/Gamma-ray detection of a nova}
The {\it Fermi} flux during the time of the {\it NuSTAR} observation is 3.1 $\pm$ 0.9 $\times$ 10$^{-10}$ erg cm$^{-2}$ s$^{-1}$ in the 0.1--300 GeV. V5855 Sgr is therefore the first nova for which we have obtained a simultaneous X-ray and gamma-ray detection.  The high N(H) requires that the shocked gas be interior to a large amount of nova ejecta.  If the ejecta are spherical and have been expanding at a constant velocity of $\sim$500 km s$^{-1}$ since the discovery of the nova (the maximum velocity observed in the absorption trough of the P Cygni profile in early spectra), then the expelled mass must be at least a few 10$^{-5}$ M$_{\odot}$ to result in such a high attenuating column on Day 13. We note that the conversion from N(H) to ejected mass is highly uncertain for two reasons.  First, it is unclear if the velocities observed in the optical correspond physically to the interacting media that are producing the X-rays; the X-rays come from a region that is behind a medium that was optically thick to visible photons at the time of the {\it NuSTAR} observation on Day 13.  Second, if the ejecta are enriched in metals (as is typically observed in novae) then the mass required to have the observed level of photoelectric absorption will be lower.  In the model fitting presented above, we assume solar abundances for our {\tt phabs} component, primarily because the statistical quality of the spectrum is too low to be sensitive to e.g. the Fe edge at $\sim$7 keV that could constrain non-solar abundances.  
%If the expansion velocity is faster than this, the mass required is higher (CHECK NUMBERS).  
 
\subsection{The majority of X-rays are thermal}
The observed spectral slope in the 10--20 keV range is not consistent those expected from known non-thermal emission processes in novae.   \citet{Vurm18} present detailed analytic models of non-thermal emission in novae that account for the observed gamma rays in novae.  The high energy emission is produced by leptons that are either accelerated directly at the shock front, or produced by $\pi^{0}$ decays.  The spectral index in the X-ray regime depends on the injection energy spectrum of the accelerated particles, $Q_{e} = dN/d\gamma \propto \gamma^{-q}$, where $q \approx 2$.  This results in a spectrum with $\nu F_{\nu} \propto \nu$ (or $\nu^{0.8}$).  Although an absorbed power law model gives a good fit to the data, the best fit value of $\alpha$ = 3.6$^{+1.3}_{-1.0}$ (or a power-law index of -3.6) implies a $\nu F_{\nu}$ spectrum that is falling with frequency. In contrast, the expected power-law index for the non-thermal emission is $-$1.0 or $-$1.2 \citep{Vurm18}. This suggests that the bulk of the X-ray emission that \nustar detected from V5855 Sgr is not associated with the low-energy tail of the gamma-ray emission.

We also considered the Compton degradation of MeV gamma rays produced by radioactive decay \citep{Livio92} as the origin for the observed X-ray emission. This mechanism was explored by \citet{Suzuki10} for the putative detection of the nova V2941~Cyg\footnote{This nova erupted a few months before the launch of {\sl Fermi\/}, so we unfortunately have no information on its GeV gamma-ray properties.} on day 9 (but not on day 29) \citep{Takei09} with the non-imaging {\sl Suzaku\/} hard X-ray detector (HXD). The HXD detection implied a very flat power law ($\alpha$ = 0.1$\pm$0.2), which was compared with Monte Carlo radiative transfer calculation by \cite{Suzuki10} starting with the $^{22}$Na decay line at 1.27 MeV and the positron annihilation line at 511 keV, the most prominent MeV features for an ONe nova. While the spectral shape was seen to be compatible with such an interpretation, the amount of $^{22}$Na required to explain the observed flux was found to be extremely high, at 3$\times$10$^{-5}$ M$_\odot$. More general models by \citet{Gomez-Gomar98}, including those appropriate for CO novae, find flat to inverted power law whose flux decays rapidly with time.  The steep spectral slope of the power-law only model for V5855 Sgr seems to disfavor this model.  Furthermore, for Compton degradation to produce observable levels of hard X-rays, the Compton optical depth must be very high. This should result in a spectrum that rises towards higher energies within the {\it NuSTAR} band. In contrast, our observations constrain $N_H$ to around $2 \times 10^{24}$ cm$^{-2}$, or Compton optical depth of $\sim$1.5. 

Given the lack of an obvious non-thermal process that produces the observed spectral slope in the 10--20 keV range, we propose that the majority of X-ray emission in V5855 Sgr is thermal in origin.  This hot plasma is produced as fast nova ejecta sweep up and shock material from a prior slower episode of mass loss.  The post-shock temperature is given by
\begin{equation}
T_{\rm sh} = 1.2~ \Big( \frac{\Delta v}{1000~{\rm km~s^{-1}}} \Big)^{2}~{\rm keV},
\end{equation}
where $\Delta v$ is the difference between the fast and slow flow velocities.  The best-fit plasma model temperature of 11$^{+11}_{-5}$ keV therefore implies $\Delta v$ $\sim$ 3000$^{+1300}_{-800}$ km/s.  The optical spectra presented in Figure 1 show H$\alpha$ P Cygni profiles on Days 3, 4 and 5 with a characteristic absorption velocity of 200--300 km s$^{-1}$.  By Day 12 the absorption wing has disappeared leaving a purely emission line with much broader widths (half width at zero intensity $\sim$3000 km s$^{-1}$).  If the P-Cygni absorption wing indicates the velocity of the slower material, then temperatures as high as 11--22 keV are difficult to account for with the maximum velocities observed on Day 12 and later.  However, the lower end of the temperature uncertainty range is reasonable.

While the bulk of the X-rays appear to be thermal, our bremsstrahlung plus power law fits indicate that the higher energy X-rays could be due to the low-energy tail of the gamma-ray emission.  In these two component model fits, the lower best fit plasma temperature of 2.2$^{+1.7}_{-0.6}$ keV implies a smaller $\Delta v$ between the fast and slow ejecta of 1350$^{+450}_{-200}$ km s$^{-1}$.  This velocity differential is smaller than that observed in the optical spectra, although we again caution that the deeply embedded location of the thermal X-ray emitting region makes it unclear if the velocities observed in the optical correspond physically to the interacting media that are producing the X-rays.

\subsection{Low thermal X-ray luminosity compared to gamma-ray luminosity}
Estimating the luminosity of any thermal X-ray emission is important for diagnosing for the properties of the shock and for assessing the efficiency of particle acceleration.  Reasonable assumptions about particle acceleration efficiency at shocks predict that a fraction ranging from 0.001 to 0.01 of the total shock power will end up being emitted as non-thermal gamma rays \citep{Metzger14}. The observed 0.1--300 GeV gamma-ray luminosity of 7.1 $\pm$ 2.1 $\times$ 10$^{35}$ ($D$/4.5 kpc) erg s$^{-1}$ would therefore lead us to expect the presence of thermal X-ray emission with luminosities in the range 10$^{37}$--10$^{38}$ erg s$^{-1}$ if all the shock power was rapidly converted to thermal X-ray emission.  However, the X-ray luminosity inferred from the best-fit single temperature plasma model is only a few 10$^{33}$--10$^{34}$ ($D$/4.5 kpc) erg s$^{-1}$.  Higher luminosities are found for the two-component models, with maximum values of a few 10$^{37}$ ($D$/4.5 kpc) erg s$^{-1}$ at the extreme of the uncertainty range.  However, these models have non-thermal X-ray luminosities that are challenging for particle acceleration models to account for (see next section for details), and the statistical evidence for this additional non-thermal component is weak.  We are then left searching for a way to reconcile the observed thermal X-ray and gamma-ray luminosities with each other.

It is possible that some Compton scattering of X-rays by electrons in the ejecta attenuates the emission beyond the photoelectric absorption level estimated by the XSPEC model. The Compton scattering cross section has no wavelength dependence, does not change the energy of the scattered photon by much in the 10--20 keV band, and is not included in the absorption models in XSPEC because the effect is very small at low N(H) and depends on geometry.  If the nova ejecta were completely spherical, Compton scattering has no net effect on the number of photons reaching us, with as many X-rays scattered into our line of sight as those being scattered out from it. However, if the ejecta are highly non-spherical this balance may by impacted and Compton scattering could be important.  At the N(H) found in our models, the Compton scattering optical depth $\tau_{C}$ is in the range 1.5--4.  If the absorbing ejecta are e.g. toroidal \citep[as inferred for several novae from optical and radio imaging, see e.g.][]{Chomiuk14}, and if we view V5855 Sgr edge-on, a significant fraction of the X-rays may be scattered out of the line of sight completely.  Correcting for photoelectric absorption only would therefore underestimate the intrinsic X-ray luminosity, by a factor of a few to 10.  It is interesting to note that this geometric effect would imply that a nova viewed pole-on would be brighter in hard X-rays at early times.  More observations of novae with {it NuSTAR} should allow us to explore this selection effect.

Alternatively, our estimate of the intrinsic X-ray luminosity may be correct, in which case we are left with the puzzle of how to account for the high ratio of gamma-ray to X-ray luminosity.  One possible explanation for the low X-ray luminosity is the suppression of thermal X-rays at turbulent, highly structured shock fronts.  Recent simulations by \citet{Steinberg18} of interacting outflows (such as those inferred to exist in novae) show that the resulting shock fronts are susceptible to thin-shell instabilities, leading to the formation of corrugated structures at the contact surface.  This results in many shock fronts being highly oblique, less efficient heating of gas, and a lowering of the X-ray luminosity by a factor of 4--36 for the parameters explored in these first simulations.  This effect may, at least in part, be responsible for lowering the X-ray emission, and could be enhanced by the non-spherical scattering losses mentioned above.

\subsection{Constraints on the gamma-ray emission mechanism}
Our spectral models have enabled us to estimate the flux (or upper limit) of any non-thermal X-ray emission in the high energy part of the NuSTAR spectra.  These limits can be compared to the measured gamma-ray flux at 100 MeV to explore the emission mechanism responsible for the non-thermal radiation.  If we assume that we have detected non-thermal X-rays above 20 keV (as in our two-component models), then the ratio $L_{X}/L_{\gamma}$ is $\sim$ 0.01 for the assumed power-law slopes of 1.0 and 1.2.  If we have not detected non-thermal X-rays, then this value is a strict upper limit.  This limit on the ratio is not very constraining, and allows for gamma rays to be produced by both leptonic and hadronic processes.  We note that achieving a ratio of 0.01 is likely unphysical for both leptonic and hadronic processes, given the high densities expected in nova shocks \citep{Vurm18}.  Electrons rapidly lose energy through Coulomb interactions in the dense nova ejecta, leading to much lower non-thermal X-ray fluxes than in the gamma-ray band.  

We noted earlier that the addition of the high energy power-law component only marginally improves the fit, although it does reduce the residual size.  We therefore do not claim to have a significant detection of non-thermal X-rays. This makes the issue of low intrinsic thermal X-ray flux even more challenging to understand.  Naively, we expect evidence of a powerful shock to show up somewhere in the emission from a gamma-ray emitting nova.  While the thin-shell instability may help with X-ray suppression and enhanced particle acceleration, the reduction is modest and cannot reduce a 10$^{38}$ erg s$^{-1}$ shock to the 10$^{33-34}$ ($D$/4.5 kpc) erg s$^{-1}$ we have detected here.  Another option is to increase the particle acceleration efficiency of these shocks.  \citet{Steinberg18} also find that the ion acceleration efficiency is enhanced at corrugated shock fronts as the magnetic field geometry is changed from a perpendicular direction to higher obliquity angles in local regions, which could result in more efficient gamma-ray production.   In combination with other effects, it is possible that the observed X-ray to gamma-ray ratio could be achieved.  

\section{Conclusions} \label{sec:results}
We have detected hard X-rays simultaneous with gamma-ray emission for the first time in a nova eruption.  From our analysis of the multiwavelength data presented here, we can conclude the following:
\begin{itemize}
\item The X-ray emitting region is deeply buried within the nova ejecta, as evidenced by the extremely high (N(H) $>$2 $\times$ 10$^{24}$ cm$^{-2}$) absorbing column found in the spectral fits.

\item The {\it bulk} of the detected X-rays are thermal in origin: the observed slope in the 1---20 keV energy range is too steep to be explained by expected non-thermal emission processes.

\item The observed thermal X-ray flux, and inferred luminosity, is much lower than expected given the observed gamma-ray flux under standard assumptions of particle acceleration efficiency in shocks. This may be due to geometric effects in correcting for absorption of X-rays, or point to intrinsic suppression of X-rays by some physical process such as thin-shell instabilities at the the internal shock front.

\item We have placed constraints on non-thermal X-rays at energies $>$20 keV, assuming the power law slopes expected for particle acceleration at a shock front.  The results from V5855 Sgr are not particularly constraining, and cannot discriminate between hadronic and leptonic scenarios.  

\end{itemize}

\acknowledgments
TN acknowledges support from the NASA {\it NuSTAR} Guest Investigator program through grant NNX17AC43G. BDM is supported through the NSF grant AST- 1615084, NASA Fermi Guest Investigator Program grants NNX16AR73G and 80NSSC17K0501; and through Hubble Space Telescope Guest Investigator Program grant HST-AR-15041.001-A.  LC is a Cottrell Scholar of the Research Corporation, and acknowledges support from NSF grant AST-1751874 and NASA Fermi Guest Investigator grant NNX14AQ36G. JLS was supported by HST-GO-13715 and Award \#2017-246 from the Heising-Simons Foundation.

%% This command is needed to show the entire author+affilation list when
%% the collaboration and author truncation commands are used.  It has to
%% go at the end of the manuscript.
%\allauthors

%% Include this line if you are using the \added, \replaced, \deleted
%% commands to see a summary list of all changes at the end of the article.
%\listofchanges

\end{document}